\documentstyle[aps,prd,epsfig]{revtex} 
\draft
\def\o{$\cal O$ }
\begin{document}
\title{%
\hbox to\hsize{\normalsize\rm May 2003
\hfil Preprint MPI-PTh/2003-23}
\vskip 36pt
         Inversion of the Photon Number Integral}
\author{L.~Stodolsky}
\address{Max-Planck-Institut f\"ur Physik 
(Werner-Heisenberg-Institut),
F\"ohringer Ring 6, 80805 M\"unchen, Germany  
~~~~~~~~~~~~~~~~~~~~~~~~~~ email: les@mppmu.mpg.de}
\maketitle
\bigskip

\bigskip

\begin{abstract} We consider the behavior of the photon number
integral under inversion, concentrating on euclidean space. The
discussion may be framed in terms of
an additive differential $I$ which arises under inversions. The
quantity $\int \int I$ is an interesting
integral invariant  
whose value  characterizes different configurations under
inversion.

\end{abstract}
\vskip2.0pc

\section{Introduction}
 
The ``photon number integral''~\cite{one}, called $n$,
is a construction whereby
one can define the number of photons radiated by a
charged particle  following a
prescribed
trajectory in space-time (minkowski space). For $n$ to be finite
the trajectory must obey  two conditions:  smoothness and
equality of 
 initial and final trajectories. The
integral then
returns a real number for any such  curve. The idea can  be
extended
from minkowksi to euclidean space where it also  becomes possible
to 
consider closed curves. The identification of initial and
final velocities (tangents) is then automatic. 

An important question  left open in~\cite{one} concerns
the  symmetries 
of  $n$. Given  $n(C)$ for a curve $C$ for what
other curves $C'$ do we have $n(C)=n(C')$? 
 There are the 
evident invariances  under rotations (lorentz transformations),
translations, and  reflections. Furthermore, since
there are no dimensional quantities except the path itself
involved in the  problem, there is
 also an evident
invariance under scaling, that is
 rescaling of all coordinates simultaneously, $x_\mu \rightarrow
\lambda x_\mu $, where $\lambda$ is a constant. 

 Nevertheless the question remains as to further invariances, in
particular with respect to conformal transformations.  These are
closely related
to scale transformations, and indeed conformal invariance
originally entered physics as 
a property of electrodynamics, and ought to be expected to apply
to
photons. Also,  in somewhat different
contexts the properties of integral expressions like
ours have been considered    
in connection with inversion 
\cite{gd}, or ``Mobius" invariance \cite{mf}, which is tantamount
to conformal invariance.
Here we would like present a discussion of 
inversions for $n$.

 The expression  in question, in minkowski space, is
\begin{equation} \label{n} 
n = \int \int dx_{\mu}{1\over S^2_{i\epsilon}}
dx'_{\mu}
\end{equation}
Since our question here is essentially a  mathematical one, we
have dropped the (dimensionless) electromagnetic coupling constant
appearing in the original expression for photons.

$S_{i\epsilon}$
is the four-distance between the points $x,x'$ in the following way
\begin{equation} \label{s} 
S^2_{i\epsilon}= (t-t'+i\epsilon)^2 -({\bf x-x'})^2 
\end{equation}

 The charm of the expression Eq~[\ref{n}] is that despite the
possible singularities it is actually finite. The possible
singularity at $S^2=0$ is handled by the $i\epsilon$ and the
possible divergence at infinity by the ``straight line condition''
for the equality of initial and final paths, as may be verified
from the fact that $n$ is zero for the simple straight
line~\cite{one}.   This means
it is well defined as it stands and needs no ``regularizations'' or
``subtractions''. In Ref~\cite{gd} a regularization is introduced
by
supersymmetry where a scalar particle cancels the singularity of
a vector particle (``gluon'') propagator.  In Ref~\cite{mf} the
singularity is cancelled 
 by  subtracting  a second  term where  the
straight line distance  $S$ between $(x,x')$ is replaced by the arc
length along the curve.  However as we anticipate from
its physical origin, Eq[\ref{n}] is finite as it stands. Indeed 
using the
identity

\begin{equation} \label{tautaua}
\partial_{\tau}\partial_{\tau'} ln(S^2)= +{4\over S^4}(\Delta
_{\mu} {dx_{\mu}\over d\tau})(\Delta _{\nu} {dx_{\nu}\over
d\tau'})- {2\over S^2} {dx_{\mu}\over d\tau}{ dx_{\mu}\over
d\tau'}\; .
\end{equation}
we were able to rewrite the integral  as
\begin{equation} \label{naab} 
 \int \int dx_{\mu}{1\over S_{i\epsilon}^2}
dx'_{\mu}=2 \int \int dx_{\mu} {\delta_{\mu \nu}- {\Delta_\mu
\Delta_\nu\over S^2 }
\over S^2} dx'_{\nu} \; .
\end{equation}
Delta is the vectorial distance between the two points, $\Delta
_{\mu}=x_{\mu}-x'_{\mu}$. We can rewrite this in a suggestive form
if we introduce
the ``transverse vector''
\begin{equation}\label{transvect}
dx^T_{\mu}=dx_{\mu}-{\Delta_\mu(\Delta\cdot dx)\over S^2 }\; ,
\end{equation}
which is $dx$  with the ``longitudinal part'' removed, i.e. has the
property $dx^T \cdot\Delta =0$. 
This is  
suggestive of the transversality property of physical photons and
it might be said that our expression is finite because it only
contains the radiated and not the ``coulomb''  photons.
In any event, the expression is now manifestly non-singular, as may
be seen by expanding the numerator, see Ref~\cite{one}.

 Observe that this procedure  does not introduce any dimensional
quantities, so
the expression is still scale invariant. Apparently the $i\epsilon$
is harmless in this respect.
This manifestly non-singular expression, without the $i\epsilon$,
may be used
to define $n$ in case of doubt and we will  use it in the following
 (mostly in its euclidean version) taking $n$ as:

\begin{equation} \label{naak} 
n= 2 \int \int dx^T_{\mu} {1\over S^2}
dx'^T_{\mu}\; .
\end{equation}

\section{Inversion}
We shall concentrate on the inversion operation

\begin{equation} \label{inv} 
x_i\to {a^2\over x^2}x_i\; .
\end{equation}
 The full conformal group is generated by adding these inversions
or ``Mobius transformations'' to the usual translations and
rotations. 
To keep the physical dimensions in order we have  introduced a
length constant $a$ parameterizing the operation. The inversion is
 around a point \o, which we  take as the origin. As we
shall
see below, the cases where the center of inversion is on the curve
$C$ itself is of special interest. The index
$i$
means any of the coordinates, and in minkowksi space includes an
inversion of the time coordinate $x_0$. We shall mainly focus
however on
the
simpler case of euclidean space.  There we 
use boldface notation for vectors, so inversion is ${\bf x}\to
{a^2\over
x^2}{\bf x}$.

We thus begin by considering the formal properties of the
euclidean expression

\begin{equation} \label{neuc} 
n= -\int \int {{\bf dx}{\bf dx'}\over S^2}=-\int \int {{\bf dx}{\bf
dx'}\over ({\bf x-x'})^2}
\end{equation}
under inversion.  The (-) sign  is the natural choice that makes n 
positive in euclidean space. The curve over which the integrations
are performed may either be a smooth closed curve, or a smooth
infinite
curve which becomes the same straight line at $\pm
\infty$.

\section{formal inversion }

 It is illuminating to start with the
relation~Eq~[\ref{tautaua}].
We use a
differential notation, where  $d_xf({\bf x,x'})=\nabla_x
f~{\bf dx}$,
so that with $S^2={(\bf x-x'})^2$
\begin{equation}\label{dlog}
d_x ln S^2={{2(\bf x-x'})~{\bf dx}\over S^2}
~~~~~~~~~~~~~~d_{x'} ln S^2=-{{2(\bf x-x'})~{\bf dx'}\over S^2}\;
.
\end{equation}
 Eq~[\ref{tautaua}]  then takes the form of  the curious  
and interesting identity
\begin{equation} \label{tautaub}
-\frac 12 \bigl[ d_x d_{x'} ln S^2 + d_x  ln S^2 ~ d_{x' }
ln S^2 \bigr] ={{\bf dx}~{\bf dx'}\over S^2}
\; .
\end{equation}
In this way we  express the integrand  of~Eq~[\ref{neuc}] in
terms
of certain 
differentials with simple transformations under
inversion. 
 Applying the inversion  to $S^2$
\begin{equation} \label{trf}
S^2=({\bf x-x'})^2\to {a^2\over x^2}{a^2\over x'^2}S^2
\end{equation}
and  so
\begin{equation} \label{lntrf}
ln S^2\to ln {a^2\over x^2}+ln{a^2\over x'^2}+ln S^2
\end{equation}
Hence if we insert the substitution Eq~[\ref{inv}]
in the lhs of Eq~[\ref{tautaub}] the first term is unchanged while
for the second

\begin{equation}\label{invdiff}
-\frac 12 \bigl[ d_x  ln S^2 ~ d_{x' }
ln S^2 \bigr]\to -\frac 12 \bigl[ d_x  ln S^2 ~ d_{x' }
ln S^2 \bigr]+
\frac 12 \bigl[-d_xlnx^2~d_{x'}lnx'^2+d_{x } ln S^2
d_{x'}lnx'^2+d_{x' }
lnS^2~d_{x}lnx^2 \bigr]
\end{equation}
In other words our fundamental form transforms additively
under inversion

\begin{equation} \label{Itrf}
{{\bf dx}~{\bf dx'}\over S^2}\to {{\bf dx}~{\bf dx'}\over
S^2}+I
\end{equation}
where we call the additional quantity 
 $I$. In this way we recover
 the results of Ref\cite{gd}
for the gluon propagator.

\section{inversion of the explicitly finite integrand }

  So far 
 we have ignored the 
singularity in Eq~[\ref{neuc}]. We return to 
 the explicitly finite 
 form 
 Eq~[\ref{naak}] as the definition of $n$. Since the
explicitly non-singular forms
 were found  (see Ref~\cite{one}) by subtracting 1/2 of
Eq~[\ref{tautaua}] from
Eq~[\ref{n}], we  do the same here and
Eq~[\ref{tautaub}] becomes

\begin{equation} \label{tautauc}
- d_x d_{x'} ln S^2 -\frac 12  d_x  ln S^2 ~ d_{x' }
ln S^2 ={2~{\bf dx}~{\bf dx'}\over S^2} -{2~[{\bf dx (x-x')}]~[{\bf
(x-x')dx'}]\over S^4}= {2~{\bf dx}^T~{\bf dx'}^T\over S^2}
\; .
\end{equation}
where  the rhs is now the explicitly non-singular integrand 
of Eq~[\ref{naak}]. The difference between the
singular and non-
singular forms is simply  the coefficient of the  first term on
the left. However, since this term
 is in any event identically invariant under inversion, 
Eq~[\ref{Itrf}] still holds, that is
\begin{equation}\label{invdiffa}
{2~{\bf dx}^T~{\bf dx'}^T\over S^2} \to {2~{\bf dx}^T~{\bf
dx'}^T\over S^2}+I\;,
\end{equation}
with  the same $I$ as in 
Eq~[\ref{Itrf}].

\section{properties of $I$}\label{iprop}
$I$ is defined as 

\begin{equation}\label{I}
I= \frac 12 \bigl[-
d_xlnx^2~d_{x'}lnx'^2+d_{x } ln S^2 d_{x'}lnx'^2+d_{x' }
lnS^2~d_{x}lnx^2\bigr]\;,
\end{equation}
where the origin is at the center of inversion \o. If the origin is
 placed elsewhere, $x$ is the distance to \o.
Evidently $I$ is symmetric
\begin{equation}\label{Isym}
I(x,x')= I(x',x)\; .
\end{equation}
Furthermore we note an important  property  arising from the fact
that
two
successive applications of the inversion is the identity operation.
Applying an inversion to the  rhs of Eq~[\ref{Itrf}]  or
Eq~[\ref{invdiffa}] again, we
should
get the original expression. We thus conclude that under inversion 
\begin{equation} \label{Isgn}
I\to -I\;,
\end{equation}
which one can also check  by  directly inserting Eq~[\ref{inv}]
into
Eq~[\ref{I}].

$I$  is a scalar under rotations, but not under translations
(holding \o fixed)
because of the presence of $x^2$, the distance from \o. It
is invariant under rescalings $\bf{x}\to \lambda {x}$ because of
the presence of the differentials. For the same reason it is
independent of the parameter $a$ giving the radius of  the
sphere of inversion. That $I$ does not contain the parameter of the
inversion suggests that it can only depend on some global
properties
of the operation.

\section{  integration}\label{prop}
To study the behavior of $n(C)$ under
inversion we proceed according to the following  steps.
 We carry out a change of variables 
according to Eq~[\ref{inv}] in the integral  for $n(C)$. This
results 
simply in an identity with a new
curve $C_{inv}$ in the new variables. The integrand receives an
additional term, $I$. 
which is also to be integrated over the new curve $C_{inv}$. That
is, we have one term with an integral of the desired expression
and one with $I$, so that
\begin{equation}\label{int}
n(C)=n(C_{inv})+\int\int\limits_{C_{inv}}I
\end{equation}
 If we can  show that $\int\int\limits_{C_{inv}} I$ is zero then
 $n(C)=n(C_{inv})$.

As another expression of the fact that  two successive operations
with
Eq~[\ref{inv}] are the identity  we can invert
$n(C_{inv})$  once more to return to the 
original curve, giving  the relation
\begin{equation}\label{inta}
\int\int\limits_{C~~~~~} I+\int\int\limits_{C_{inv}} I=0\; .
\end{equation}
This of course just amounts to Eq~[\ref{Isgn}] if we change
variables
again to make ${C_{inv}}\to{C}$.
An evident property following from
 Eq~[\ref{inta}] is that if ${C_{inv}}\equiv{C}$, as for a
circle   with the center of inversion in the center, or
a closed curve  
in 3-space  on the surface of the sphere
of inversion, then

\begin{equation}\label{equiv}
\int\int\limits_{C~~~~~} I=0~~~~~~~~~~~~~~~~~~C\equiv{C_{inv}} \;
.
\end{equation}

  As opposed
to Eq~[\ref{Isgn}] 
this is not merely an algebraic identity but involves of course the
nature of the curves.
It is evidently true by Eq~[\ref{Isgn}]  if $C$ and $C_{inv}$ are
identically the same
curve. But it is clearly also true in some more general sense, say
if  one
curve is simply a rotation, translation or rescaling of the other.
The wider meaning of ``$\equiv$'' in $C\equiv{C_{inv}}$ is an 
interesting question and
will be
further discussed below (section \ref{sum}).

\section{ two curves}
An amusing generalization of these properties of the integration
suggests
itself. We mention it although it lies somewhat outside our main
topic. The quantity $\int\int I$ is a functional of one curve
$C$. But  actually it could be regarded as a functional of two
curves
$C$ and $C'$. We might
have {\it different} curves in the $x$ and $x'$ spaces. That is,
we consider the integral $\int_C\int_{C'}I$.

 Because  of Eq~[\ref{Isym}] it doesn't matter to which variable
the curve is assigned:$\int_C\int_{C'}I=\int_{C'}\int_{C}I$. Now
let $C'$ be the inversion of $C$ so $C'=C_{inv}$. Then under
inversion 
 according to Eq~[\ref{Isgn}] the integral goes to minus itself.
Therefore

\begin{equation} \label{idf}
  \int_C\int_{C'}I=0~~~~~~~~~~~~~~~~~~~~~C'=C_{inv}\;.
\end{equation}
Eq~[\ref{equiv}] may then be read as the special case  where
$C=C_{inv}$.

A place where Eq~[\ref{idf}]  might be useful concerns the question
of the additivity of the parts of a curve. In general of course 
 $\int\int I$ for a single curve cannot broken into two parts
 such that the total integral is the sum of the integrals for the
two parts. 
$I$ is a bilocal object, and there will usually be cross terms 
between the two parts. However, if the two parts in question are
the inversion of each  other, 
then according to Eq~[\ref{idf}], the cross terms are zero and the
integrals for the two parts may be simply added.

To develop this idea properly a study of the possible
singularities,
which may be different than in the single-curve case, would be
necessary.
 We stress that  our
integrals are always meant over just one
curve  in the original sense unless explicitly indicated as in
Eq~[\ref{idf}].

\section{ Possible Singularities of $I$}

 There are singularities or potential singularities of $I$ which
need to be understood.  We will come to the conclusion that,
despite appearances, $I$ is a non-singular object--- for a given
curve. 

We would like to argue that  for the investigation of possible
singularities at a point  it suffices to consider the
behavior of $I$ for a straight line through that point.
 Consider the smooth curve ${\bf x}(\tau)$ in the neighborhood of
the
point ${\bf x}(0)$. It is convenient to introduce the ``natural''
parameterization where $\tau$ is the length along the curve 
(proper time, in minkowski space). Then ${\bf x}\approx
{\bf x}(0)+ \dot {\bf x}\tau+\ddot {\bf x}(1/2)\tau^2+..$, where
$\dot{\bf x}, \ddot{\bf x},...$ are the first, second,...
derivatives
with respect to $\tau$. Due to the choice of $\tau$ as the path
length $(\dot{\bf x})^2=1$, $\dot {\bf x}\ddot{\bf x}=0,$ and so
forth. Now
${\bf x}(\tau)-{\bf x}(\tau')$ is odd with respect to interchange
of $\tau,\tau'$.  This
leads to $({\bf x}(\tau)-{\bf x}(\tau'))^2 \approx (\tau-
\tau')^2\bigl[(\dot{\bf x})^2 +(\ddot{\bf x})^2
b(\tau,\tau')+...\bigr] $, 
where $b$ is a bilinear  expression in $\tau,\tau'$.

Since, with $(\dot{\bf x})^2=1$ 
\begin{equation}\label{lnt}
ln\bigl[(\tau-\tau')^2(\dot{\bf x}^2+(\ddot{\bf x})^2
b(\tau,\tau')+...)\bigr]=ln(\tau-\tau')^2+ln \bigl[1+(\ddot{\bf
x})^2 b(\tau,\tau')+...\bigr]
\end{equation}
we see that the possibly singular behavior of the logarithm results
from the ``velocity'' or tangent term $\dot {\bf x}$, while the
dependence on the ``acceleration'' or curvature is non-singular for
$\tau, \tau' \to 0$.
Similarly for $ln x^2$ near $x^2=0$ we have $ln x^2\approx
ln \tau^2+ln(1+(\ddot{\bf x})^2 b(\tau)+...)$.    
We may thus investigate the possible singularities  by
looking at the behavior of $I$ for a straight line.

We first consider the possible singularity for $S^2=0$,
 while  assuming $x^2\neq0$, that the center of inversion  \o is
not
on the curve.  

\begin{figure}[h]
\epsfig{file=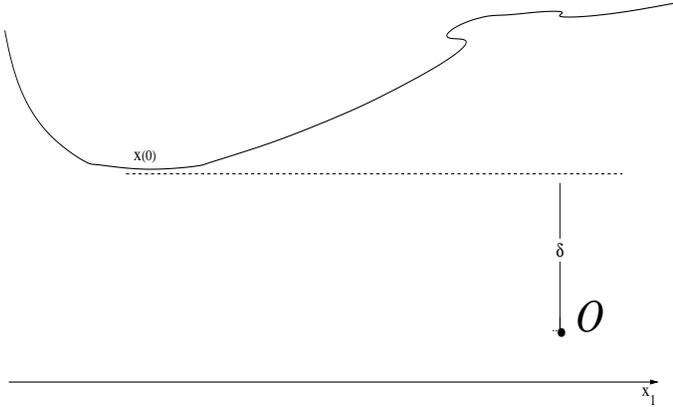, height=.3\hsize, width=.5\hsize}
\caption{Construction for studying  the possible singularity due to
$S^2=0$ at a point $x(0)$. The direction of the tangent (dotted
line) is used to determine the ``1'' axis and has ``impact
parameter'' $\delta$ with respect to the center of inversion. }
\end{figure} Choose the ``1'' axis such 
that it is parallel to the tangent $\dot {\bf x}$ at the
point. Let the distance of the projected tangent from \o i.e. its
``impact parameter'', be $\delta$ (Fig 1). Now, for
the investigation of  the possible singularity we treat the
curve as this straight line, as just explained. We then have
$S^2 = ({x_1}- {x'_1})^2$
and  $d{\bf x}=dx_1$. 
After some algebra following  the manipulations of Ref~\cite{gd}
for
the straight line we have
\begin{equation}\label{Ist} 
 I_{straight ~line}=-2~dx_1 dx'_1{\delta^2\over
{(x_1}^2+\delta^2)
( {x'_1}^2+\delta^2)}\;, 
\end{equation}
which is in fact non-singular at ${\bf x}\approx {\bf x'}$.  The
absence of
the singularity may be traced to the fact that $I$ is
symmetric in
$x,x'$ and so must contain only even powers in $\bf x-x'$. The
potential
singularity  from Eq~[\ref{dlog}], however would be odd in this
variable, and so is in fact absent.

Concerning the possible singularity at $x^2=0$, as occurs for a
curve through \o, first consider 
$I(0,x')$, i.e.  one variable at some ordinary point and the
other near zero. In this case the first and third term of $I$,
Eq~[\ref{I}] cancel since $d_{x'}ln S^2=d_{x'}ln x'^2$ for $x=0$.
On the other hand the second term of $I$ is non-singular, so there
is no singularity.

As for $I(0,0)$, we examine the straight line through the origin.
By direct inspection of $I$ one finds

\begin{equation}\label{Ist0} 
 I_{straight ~line~through~{\cal O} }=0\;. 
\end{equation}
Therefore  there are no singularities connected with $x=0$.

Eq~[\ref{Ist0}] may be viewed as the consequence of a general
symmetry property since two points along a ray may be interchanged
by an inversion. Therefore by Eq~[\ref{Isgn}] $I(x,x')=-I(x',x)$.
On the other hand, by Eq~[\ref{Isym}] we also have 
$I(x,x')=+I(x',x)$, so $I$ is zero.

Since in Eq~[\ref{Ist0}]  we have zero for the straight line
through
\o, we look at
the simplest
curvature or ``acceleration'' contribution,
that of  the circle.  
Consider a circle passing through \o and use the relation for the
length of a chord $l=2Rsin\phi$, where  $\phi$ is  the
half-angle subtended at the center of the circle. This leads to

\begin{equation}
I=1/2\bigl[-d\,ln\,sin^2\phi~ d\,ln\,sin^2\phi'+d\,ln\,sin^2(\phi-
\phi')~d\,ln\,sin^2\phi'+d\,ln\,sin^2(\phi-\phi')
~d\,ln\,sin^2\phi\bigr]
\end{equation}
that is $I=2d\phi d\phi'\bigl[-cot\phi~cot\phi'+cot(\phi-\phi')(
cot\phi'-cot\phi)\bigr]$.
Using the identity $cot(\phi-\phi')={cot\phi cot\phi'+1\over
cot\phi'-cot\phi}$ we have finally

\begin{equation}\label{circleoa}
I_{circle~through~{\cal O} }=2d\phi d\phi'\;,
\end{equation}
where $0\geq\phi\geq\pi$.

If we have only a  segment of a circle passing through \o the
integration will only be over the corresponding angle, and
Eq~[\ref{circleoa}] may be said to approach Eq~[\ref{Ist0}] in the
sense that as the radius of the  circle $R$ becomes very large this
angular segment becomes very small. 
The very simple form of Eq~[\ref{circleoa}] suggests that it may
sometimes be preferable to use coordinates not centered
on \o but rather on the center of curvature for the curve at \o.

\section{simplest or ``reference'' cases}
 There are four configurations to discuss, according to whether the
curve is closed or infinite and whether
\o is on or off the curve. For orientation we discuss a
simplest or ``reference'' curve for each case.

\subsection{circle, not through \o}\label{cnot}

This is the simplest case of the finite closed curve where the
center of inversion is not on the curve. Inversion of a circle
produces another circle, and similarly inversion of a general
closed curve will produce another closed curve, as long as \o is
not on the curve.

 For the circle one finds $n$  by directly integrating 
Eq~[\ref{naak}]

\begin{equation}\label{circle}
n_{circle}=2\pi^2 \; .
\end{equation}
 Since $n$ is purely a property of the geometric figure, this holds
for any circle.

 We therefore conclude from Eq~[\ref{int}]

\begin{equation} \label{ncirco}
\int\int I_{circle~not ~through~\cal O}= 0 \;.
\end{equation}

 As on check on this argument we can show directly  that the
integral of $I$ is zero by Eq~[\ref{equiv}], that is by
Eq~[\ref{Isgn}], for those  circles going identically into
themselves under inversion. This occurs for inversion through the
center of a circle for example, or when \o is in the plane but
outside the circle,
 take $a^2=dD$ in  Eq~[\ref{inv}]  with
$d$ the closest
point of the circle to \o and  $D$ the furthest.

\subsection{ infinite straight line not through \o, or circle
through \o }

We now turn to the consideration of curves involving infinities.
The simplest case is the infinite straight line, not through \o.
Under inversion this becomes a circle through \o. The general case
here  refers to curves, which
although they become the same  straight line~\cite{line} at large
distances,
have some arbitrary
form at finite distances (Fig 3). As with the straight line/circle
pair,  under inversion these become a closed curve passing through
\o; and vice-versa.
 Since they are the images of each other under inversion,
we consider the two  cases together.

That some subtlety is involved  is evident from the fact
that on the one hand we have 
\begin{equation}\label{stline}
n_{straight~line}=0 \; .
\end{equation}
But for the inversion of
the
straight line,  a circle crossing the origin (or for that matter
any
circle),  we  have  not zero but rather  
\begin{equation}\label{circlea}
n_{circle}=2\pi^2 \; ,
\end{equation}
as is found by  integrating the definition Eq~[\ref{naak}].

That different values for $n$ result is perhaps not entirely
surprising
since the inversion has produced a basic  change in the figure, an
infinite curve becoming a finite curve. In Ref~\cite{gd} this
difference was associated with the
``anomaly''.

 According to Eq~[\ref{int}],
 when starting  from the straight line we must
integrate $I$ over
a  circle through \o. Or  when beginning with a
circle we
 integrate $I$ along an infinite straight line. To be consistent
with Eq~[\ref{stline}] and Eq~[\ref{circle}], these
integrations  ought
to produce  non-zero and opposite sign  contributions. 

Indeed, in Eq~[\ref{Ist}] we already have 
$I$ for the straight
line. Carrying out the integral we see that it is independent of
$\delta$ and gives
\begin{equation} \label{nst}
\int\int\limits_{-\infty~~~~}^{+\infty~~~~}
I_{straight~line~not~through~\cal O }= -2 \pi^2\;
.
\end{equation}

Similarly, we  use Eq~[\ref{circleoa}] to integrate
over a circle through \o

\begin{equation} \label{ncirc}
\int\int I_{circle~through~\cal O}= 2 \pi^2\;
.
\end{equation}

These results are in agreement with Eq[\ref{int}], 
Eq~[\ref{stline}] and
Eq~[\ref{circlea}]
and of
course Ref~\cite{gd}.
We note the contrast of Eq~[\ref{ncirc}] with Eq~[\ref{ncirco}]; or
between  Eq~[\ref{nst}] and Eq~[\ref{intsto}] below. Apparently
while $I$
is a non-singular object along a {\it given} curve, an
infinitesimal
change in that curve  can produce a finite effect in the integral.

\subsection{ straight line through \o}
For closed finite curves, as typified by the circle not through \o,
we had neither points at infinity nor
points at zero. In the previous subsection we had either one or the
other, mapping into each other under inversion. Finally, we 
consider  having both at the same time:  curves which are both
infinite and  pass
through \o. The simplest or reference case is the straight line
through
\o. An inversion returns the same straight line
through the \o. Observe that also for the general case of this
class, namely an 
arbitrary curve passing through \o and 
becoming  the same straight line at $\pm\infty$,  inversion
produces a curve of this same type (Fig 2).

 In Eq~[\ref{Ist0}] we had
$I_{straight ~line~through~ {\cal O} }=0$. Evidently then
\begin{equation}\label{intsto}  
\int\int I_{straight ~line~through~ {\cal O }}=0\;.
\end{equation}

This is of course in agreement with the arguments of section
\ref{prop} since under  inversion the straight line through \o,
being on a  ray, goes
identically into itself.

\begin{figure}[h]
\epsfig{file=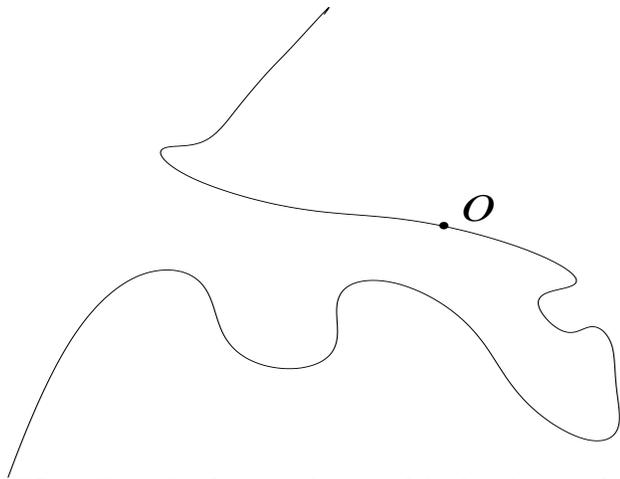, height=.35\hsize, width=.5\hsize}
\caption{Example of a general curve of the fourth type of
configuration
``straight line through \o ''. The lines continue to $\pm \infty$
along a common straight line, and  an arbitrary number of euclidean
dimensions is implied.}
\end{figure}

This case is most liable to be of interest physically  in
minkowski space where it will correspond to the path of a charge,
and the 
inversion also produces a possible path for a charge.

\section{once-integrated expression}\label{oie}
We now turn to the explicit integration of $I$. 
To integrate
 $I$, note that it consists of product pairs where  one member
of each product does not contain both variables. 
Hence one member of each pair, given as a total derivative, can be
explicitly integrated.
 The limits of integration are
also the same in both variables. Therefore it appears useful to
interchange the names of $x,x'$ say in the last term of $I$,
Eq~[\ref{I}] to
arrive
at the expression $(1/2)\int\int d_x[-lnx^2+2ln{(\bf x-x')}
^2]d\,lnx'^2$, where the  $x$
integration can be
done immediately. However, this has the disadvantage that it is no
longer explicitly symmetric in $x,x'$, which introduces a
singularity of the integrand for ${\bf x}\approx {\bf x'}$.
Although this is only
apparent and integrates to zero, it is perhaps more comfortable  to
let
$ln [S^2]\to ln [S^2+\eta^2]$ so as to make the
integrand 
 explicitly finite and then let $\eta\to 0$ at the end of the
calculation.
It may be verified that $I$ is completely regular under this
procedure.  
 After the relabeling of variables we have
an explicitly finite expression (for \o not on the curve)
and can now perform the $x$ integration along an arbitrary  curve
between  limits $\bf A$ and $\bf B$ to obtain

\begin{equation}\label{ab} 
\int \int\limits_{\bf B~~~~~}^{\bf A~~~~~} I= \int_{\bf B}^{\bf
A}d\,ln x^2
\biggl( -
\frac 12\,ln{A^2\over B^2}+ln{({\bf x-A})^2\over({\bf x-B})^2
}\biggr)~~~~~~~~~~~~~~ $C$~ not~
through~ {\cal O}\;.
\end{equation}

 Observe that there are no singularities connected with this
expression. At an endpoint, for $\bf x\to A$ 
we  have $\int^{\bf A} d\,lnx^2 ln({\bf x-A})^2\sim \int^1
dx~(lnA^2
+ln(x-1)) $ which is non-singular; while $x\to 0$  is excluded by
assumption.

However, we shall also need a once-integrated expression when  \o
is on the curve, where $x\to 0$ must be considered.  To see what
replaces Eq~[\ref{ab}] note that
 Eq~[\ref{circleoa}] or Eq~[\ref{Ist0}]  tell us that there is in
fact no singularity associated with \o. Therefore  we
can excise an
infinitesimal region along the curve
around $x=0$  without affecting the value of the integral: 
\begin{equation} \label{ex}
\int\int\limits_{\bf B~~~}^{\bf A~~~} 
I=\biggl(\int_{\bf B}^{\bf- \epsilon}+\int_{\bf \epsilon}^{\bf
 A}\biggr)\biggl(\int_{\bf B}^{\bf- \epsilon}+\int_{\bf
\epsilon}^{\bf A}\biggr) I \; ,~~~~~~~~~~~~~~~~~~~\epsilon \to 0
\end{equation}

 Now again  writing $\int\int I=(1/2)\int\int d_x[-lnx^2+2ln{(\bf
x-x')}
^2]d\,lnx'^2$ and carrying out the $\int d_x$ we obtain

  \begin{equation}\label{abon}
\int\int\limits_{\bf B~~~~}^{\bf A~~~~} I_{\,C~through~\cal
O}=\biggl(\int_{\bf
B}^{\bf -
\epsilon}+\int_{\bf \epsilon}^{\bf A}\biggr) d\,ln x^2  \biggl(
-\frac 12 ln
{A^2\over B^2}+ln{({\bf x-A})^2\over({\bf
x-B})^2}+ln{({\bf x+ \epsilon})^2\over({\bf x-
\epsilon })^2}\biggr) \; .
\end{equation}

 We first consider the last term, $ \int_{\bf B}^{\bf -
\epsilon}+\int_{\bf \epsilon}^{\bf A} dln x^2\, ln {({\bf
x+\epsilon})^2\over({\bf
x-\epsilon })^2}$. As $\epsilon \to 0$, the integrand
vanishes for any non-infinitesimal value of $x$. Thus it suffices
to evaluate the integral for a straight line in the vicinity of \o.
Introducing the variable $y=x/\epsilon$, this  leads to the
integral $\int_{-\infty}^{-1}+\int^{\infty}_{1} {2 dy~\over y}  ln
({y+1\over
y-1})^2=8 \int^{\infty}_{1}{dy~\over y} ln ({y+1\over
y-1})=2\pi^2$~\cite{eduardo}, so we may write

\begin{equation}\label{lim}
\int_{\bf B}^{\bf -
\epsilon}+\int_{\bf \epsilon}^{\bf A} d\,ln x^2 
\, ln{({\bf x+\epsilon})^2\over({\bf x-\epsilon
})^2}=2\pi^2~~~~~~~~~~~~~~~~\epsilon \to 0 \; ,
\end{equation}
for any smooth curve through \o, with {\bf A} and {\bf B}
on opposite sides of \o (compare Eq~[2.15] of Ref\cite{gd}).

As for the remaining part of Eq~[\ref{abon}], we verify that it is
non-singular. For the endpoints, say $\bf x\to A$, it is finite 
for the same reason given above with regard to Eq~[\ref{ab}]. For
$x \to 0$, we write $ln{({\bf x- A})^2\over({\bf x-
B})^2}= ln {A^2\over B^2}+
 ln (1-2{{\bf Ax}\over A^2}+{x^2\over A^2})- ln (1-2{{\bf Bx}\over
B^2}+{x^2\over B^2} )$. The constant terms  $\int_{\bf B}^{\bf
-\epsilon}+\int_{\bf \epsilon}^{\bf 
 A} (1/2)ln{A^2\over B^2}=(1/2)ln{A^2\over B^2}(ln{\epsilon^2\over
B^2}+ln{A^2\over\epsilon^2 })=(1/2)(ln{A^2\over B^2})^2$ are
nonsingular.  For the $x$-dependent terms we can write $ln
(1-2{{\bf
Ax}\over A^2}+{x^2\over A^2})\approx -2{{\bf Ax}\over A^2}$, and
similarly for the $B$ term, leading to an integral of the type
$\int x\, d\,ln x$ which is also  non-singular at $x=0$.

Thus the expression is $\epsilon$ independent and  well defined,
and
we introduce the symbol $\cal P$ for this  principal value-like
integral: $\cal P \int_{\bf B}^{\bf A}=\int_{\bf B}^{\bf -\epsilon}
+\int_{\bf \epsilon}^{\bf A}$ and can thus finally write for
Eq~[\ref{abon}]

\begin{equation}\label{abona}
\int_{~\bf B}^{~\bf A}\int I= 2\pi^2+{\cal P} \int_{\bf B}^{\bf A}
d\,ln x^2  \biggl(
- \frac 12\, ln
{A^2\over B^2}+ln{({\bf x-A})^2\over({\bf x-
B})^2} \biggr)~~~~~~~~~~~~~~~C~through~\cal O \;.
\end{equation}

We wish to use Eq~[\ref{ab}] and Eq~[\ref{abona}]
in the next section to extend the  simple results found for
straight lines
and circles to general curves, but first we check that these
formulas give the expected results in these simple cases:

1)For the circle, not through \o, we expect zero, according to
Eq~[\ref{ncirco}], which is indeed what results from
setting $\bf A=B$ in Eq~[\ref{ab}].

2)For the infinite straight line, not through \o, we expect
$-2\pi^2$
according to Eq~[\ref{nst}]. Using Eq~[\ref{ab}], we consider the
straight line at a distance $\delta$ from \o, as in Eq~[\ref{Ist}]
and  introduce the variable $y=x/\sqrt{AB}$. In the limit $A, B\to
\infty$ such that $A/B \to 1$ and $\delta/\sqrt{AB} \to 0$ one
obtains the integral $4\int_{-1}^1 {dy\over y} \; ln{1-y\over 1+y}
$, which is  indeed $-2\pi^2$.

 3)Turning now to \o on the curve and  Eq~[\ref{abona}], for the
circle we expect $2\pi^2$, according to Eq~[\ref{ncirc}]. This is
what we obtain upon setting $\bf A=B$ in Eq~[\ref{abona}].
   
4)For the final  example, the straight line through \o, we
expect zero according to Eq~[\ref{intsto}]. To evaluate the
principal value integral, we repeat the arguments just given for
the infinite straight line not through \o, with the difference
that in place of $\delta$, we now
have $\epsilon /\sqrt{AB} \to 0$. This leads to  $8\int_{0}^1
{dy\over y} \; ln{1-y\over 1+y} =-2\pi^2$ again and 
  Eq~[\ref{abona}] is zero.

We emphasize  that the complications of this section arise from our
desire to bring $\int \int I$ into the once-integrated form and the
resulting asymmetric treatment of the variables; $I$ itself is
perfectly well behaved  for a given curve. Perhaps if another
method could be found for the problem of  general curves as
discussed in the next section these
complications could be avoided.

\section{Integral of $I$ as an invariant}
 Perhaps the most remarkable property of $\int \int I$ is that it
is a type of invariant, having the same value for all curves of a
given configuration.  In our language,  this was what was
essentially concluded
in Ref~\cite{gd}   for certain field theoretic   amplitudes (for
our
cases 1-3). We shall show this  using 
Eq~[\ref{ab}] and Eq~[\ref{abona}] of the previous section.
Since the quantity in parenthesis in Eq~[\ref{ab}] and
Eq~[\ref{abona}] does not depend on the
particular curve, the nature of the actual curve in question
appears only in the remaining single integration over $x$. This
feature that makes the once-integrated expressions useful for the
examination of general curves. 

\subsection{ closed  curves not through
\o}\label{seceuc}
For our first  case we take the generalization for
section~\ref{cnot} : inversion of a general,
 finite closed euclidean curve, with \o not on the
curve.  The inversion operation
produces
another finite
closed curve. Therefore we can set $\bf A=B$ and
the integrand of Eq~[\ref{ab}] is zero. 

\begin{equation} \label{intf}
\int\int I_{closed~not~through~ \cal O} =0\; .
\end{equation}
This is naturally in agreement with the example Eq~[\ref{ncirco}],
and in view of Eq~[\ref{int}] we can finally conclude that for this
case $n$ is inversion invariant:

\begin{equation}\label{neq}
n(C)=n(C_{inv})
\end{equation}
where $C_{inv}$ is the inversion of any finite closed
euclidean curve $C$, with the center of inversion not on the curve
$C$.

\subsection{ infinite curves not through \o }\label{gif}
 We now consider the generalization of the infinite  straight line.
By these we mean    curves which
although they become the (same)  straight line at large distances,
have some
arbitrary
form at finite distances (Fig 3). 

We employ the once-integrated 
Eq~[\ref{ab}] again, where the  limits {\bf A} and {\bf B}
limits are to be sent to $\pm \infty$.
The quantity in  parenthesis in Eq~[\ref{ab}] would be the same for
any curve and in
particular for the straight line. Since the curve under
consideration only differs from the straight line in a finite
region, say between points $\bf F$ and $\bf G$, let us add and
subtract 
a straight line contribution from $\bf F$ to $\bf G$  in
Eq~[\ref{ab}]. This leads to

\begin{equation}\label{abc}
\int_{\bf B}^{\bf A}d\,ln x^2 (...)=\int_{straight~line }d\,ln x^2
(...)+\int_{finite}d\,ln x^2(...)\; .
\end{equation}
The $(...)$ stands for the parenthesis in Eq~[\ref{ab}].
$\int_{straight~line }$ is from {\bf B} to {\bf A},  and  as  {\bf
A, B}  are sent to $\pm \infty$ it becomes the integral
 for
the infinite straight line, $-2\pi^2$. The
second
term, $\int_{finite}$, stands for the integral from $\bf F$ to $\bf
G$ along the actual curve in question minus the integral $\bf F$ to
$\bf G$ along a straight line.

 We  now  argue that $\int_{finite}$ goes to zero  as 
${\bf A},  {\bf B}\to \pm \infty$.  This is because  $x$ in
$\int_{finite}$ is confined to finite values as say ${\bf A}\to
\infty$. Thus we can write $ln\bigl[({\bf x-A})^2\bigr]= ln
A^2 +ln\bigl[1-2{{\bf Ax}\over A^2}+{x^2\over A^2} \bigr]\approx ln
A^2 -
2{{\bf Ax}\over A^2} $ to leading order in $1/A$. The $2{{\bf
Ax}\over A^2} $ will lead to a contribution to the integral
vanishing as $1/A$. Then $\int_{finite}$ becomes the difference of
the integral of a
total derivative over two paths between the same endpoints  and so
is zero.

   ( There might appear to be some difficulty with the argument
when \o is located such that the straight line from 
$\bf F$ to $\bf G$ passes through it, i.e when the straight lines
at infinity lie along a ray from \o. We can deal with  this
by replacing the straight line as the reference curve by a curve
 where a semi-circle avoids \o. In this case the semi-
circle contribution goes to zero as ${\bf A}\to \infty$, while the
straight line integrals can be evaluated in this limit to again
give $-2\pi^2$.)

 We thus conclude that the integral $\int\int I$ over any curve,
not passing through \o,
differing from the straight line in a finite region is the same 
as that for the simple straight line, not passing through \o:
\begin{equation} \label{intinf}
\int\int I_{infinite~not~through~ \cal O} =-2\pi^2\; .
\end{equation}

\begin{figure}[h]
\epsfig{file=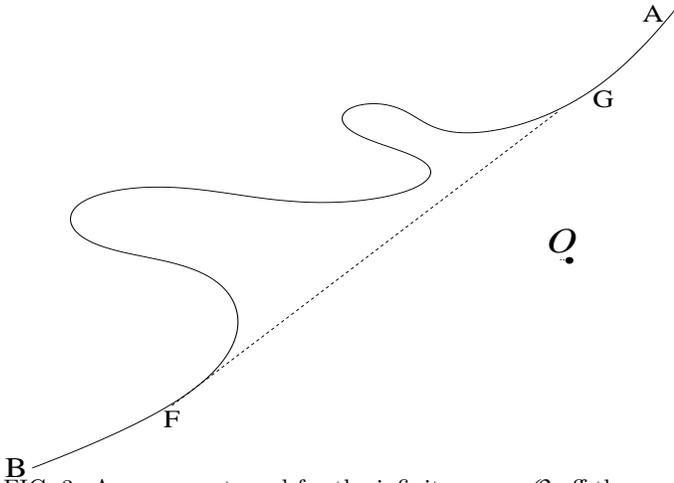, height=.35\hsize, width=.5\hsize}
\caption{Arrangement used for the infinite curve, \o off the
curve. The sketch is to be understood in an arbitrary number of
euclidean dimensions. The solid line represents the curve under
consideration. The contribution from the dashed line is to be added
and subtracted so as to produce the integral for the infinite
straight line  as ${\bf
A}, {\bf B}\to \infty$ plus a contribution between $\bf F$ and $\bf
G$ which contains  only finite values of $x$.}
\end{figure}

\subsection{closed curves through \o}
Inversion now leads to
the generalization of the circle through \o , the arbitrary closed
curve through \o. We use  Eq~[\ref{abona}], and setting  {\bf A}=
{\bf B},  $\int \int I$ is the same as for
the circle:
\begin{equation} \label{intcirc}
\int\int I_{closed~through~ \cal O} =2\pi^2\; .
\end{equation}
 And for $n$ we can say

\begin{equation}\label{gnf} 
n(C)=n(C_{inv}) +2\pi^2\; .
\end{equation}
Where $C$ is a generalized circle and $C_{inv}$ is its inversion,
a generalized infinite straight line.

\subsection{infinite curve through \o}

We come to our last case, the generalization of the straight line
through \o. By this we mean an infinite curve with our usual
condition that it becomes the same straight line at infinity, but 
now
also passing through \o at some finite point (Fig 2). This class
inverts into itself.

We now use 
Eq~[\ref{abona}] where we must evaluate $ {\cal P} \int_{\bf
B}^{\bf A}d\,ln x^2 (...)$.  In this integral, adding and
subtracting a straight line piece between  the finite points ${\bf
F}$ and ${\bf G}$ (As in Fig 3 but with \o on the solid curve), we
have again

\begin{equation} \label{pon}
 {\cal P} \int_{\bf B}^{\bf A}d\,ln x^2 (...)= \int_{straight~line
}d\,ln x^2
(...)+{\cal P}\int_{finite}d\,ln x^2(...)\; ,
\end{equation}
where we have dropped the $\cal P$ in $\int_{straight~line }$ since
it does not pass through \o. Since $\int_{straight~line }=-2 \pi^2$
as $\bf A,B \to \infty$, the $2 \pi^2$ from Eq~[\ref{abona}] is
cancelled and we are left with
\begin{equation} \label{pona}
{\cal P}\int_{finite}d\,ln x^2(...)=\bigl(-\int_{\bf F}^{\bf G}
+{\cal P} \int_{\bf F}^{\bf G}\bigr)d\,ln x^2\biggl( -
\frac 12\,ln{A^2\over B^2}+ln{({\bf x-A})^2\over({\bf x-B})^2
}\biggr) \; ,
\end{equation}
where the first integral is along the straight line piece and the
second integral is along the curve in question.

 As before, for large  $\bf A,B $ the $(...)$ goes  to an $x$
independent piece and terms vanishing as  $\bf A,B \to \infty$.
Since $\int_{\bf F}^{\bf G}d\,ln x^2={\cal P} \int_{\bf F}^{\bf
G}d\,ln x^2 $ for the two different paths, Eq~[\ref{pona}] goes to
zero and we can conclude

\begin{equation} \label{intinfo}
\int\int I_{infinite~through~ \cal O} =0\; 
\end{equation}
for arbitrary curves. This implies finally for $n$  

\begin{equation}\label{neqa}
n(C)=n(C_{inv})
\end{equation}
where $C_{inv}$ is the inversion of an infinite 
euclidean curve $C$, with the center of inversion  on the curve.

\subsection{summary}\label{sum}

We can summarize the results of this section as follows. 
We have two types of curves: ``finite'' curves and ``infinite''
curves. The first are closed curves, the second open curves 
becoming the same straight line  at large distances. There are also
two possibilities for the placement of the origin of inversion \o:
``on'' and ``off'' the curve. Labeling the configurations from 1 to
4
we can  exhibit their properties   in a table:  
\begin{center}
\vskip.5cm
\begin{tabular}{|l|l|l|l|l|} 
\hline
CONFIGURATION&CURVE~~~~~~~ & $\cal O$~~~~~~~ &RESULT~~~~~~~~~
&$\int\int
I$~~~~~~~~\\
\hline
\hline
1&finite&off&1&$0$\\
\hline
2&finite&on&3&$2\pi^2$\\
\hline
3&infinite&off&2&-$2\pi^2$\\
\hline
4&infinite&on&4&0\\
\hline
\end{tabular}
\end{center}
\vskip.5cm

The  column ``Result'' refers to the configuration resulting
from
the inversion. Thus: configuration 2 is a closed finite curve with
\o on the curve. The integral of $I$ over this curve has the value
$2\pi^2$.
The inversion produces configuration 3, which is an infinite curve
with \o off the curve.

A notable feature of the table is that
 those configurations, namely 1 and 4, which invert into themselves
have $\int\int I=0$. This is  a type of generalization of 
Eq~[\ref{equiv}], and appears to answer the question raised in
sect \ref{prop} as to the wider meaning of ``$\equiv$''. Apparently
two curves should be considered ``equivalent'' when they belong to
the same configuration  in the sense of
the table. Depending on the position of \o, a curve may map into
its own configuration or not. When it does, the ``self conjugate''
property $\int\int I=0$ obtains.

\section{ infrared/ultraviolet duality}
  We remarked in
\cite{one} that it appeared as if the finiteness of $n$ both at
short distances (ultraviolet) and at long distances (infrared)
could in a sense be attributed to the same thing, namely that the
average velocity $U(x,x')$ and the instantaneous velocity $u$
become
equal.
In euclidean space $U$ and $u$ refer to the chord and tangent of
the
curve respectively.
At short distances as $x\to x'$we  have $U \to u$ because the curve
is taken to be
smooth,
so for small enough intervals the chord and the tangent become the
same. At large
distances  the curve has  by assumption a constant and equal  slope
at $\pm \infty$, so that $U \approx u$ again.

 It is interesting to note how the   configurations of the table
where \o is ``on'', that is where ``Result'' is an infinite curve,
represents just this 
situation. 
 In  inversion
 a point on the curve is projected along the
(directed) ray connecting
it to the origin. The point  goes to large distances if it was
close to
the origin and goes to small distances if it was far
from the origin. If \o is directly on the
curve,  the points approaching the origin along the curve from one
side will be
sent to a straight line at $+\infty$, while points approaching
\o from the other side will be sent to a straight line at $-\infty$
in the opposite direction. That  the {\it same, straight} line at
$\pm\infty$ results is a consequence of the presumed smoothness of
the curve at \o.

 Similarly, a curve coming from  large distances and finally going
to large distances as the same straight line  will be mapped into
a curve
smooth at the origin. If the infinite slopes had been different 
there would be a kink at the origin.
 
 Inversion  thus makes it clear how our two assumptions needed to
make $n$ finite both in  the infrared and the ultraviolet are
related: under an inversion whose origin is on the curve, curves
which are locally smooth  become straight and parallel at $\infty$
and vice-versa.

\section{minkowski space}

 In minkowksi space a new aspect enters in that Eq~[\ref{inv}] can
make points that are time-like separated into points that are
space-like separated. That is,
 in  minkowski space, where $S^2=(x_0-x_0')^2-({\bf x-x'})^2$ the
purely algebraic relation Eq~[\ref{trf}] is still true

  \begin{equation} \label{trfa}
S^2 \to {a^2\over x^2}{a^2\over x'^2}S^2\; ,
\end{equation}
but for an initially physical time-like curve, with \o ``off''
there
will be points on it space-like to \o . This  gives $x^2$ negative
in Eq~[\ref{trfa}] and a  time like-like separation
$S^2>0$ may be mapped into a space-like separation $S^2<0$
~\cite{pairs}.

On the other hand for a time-like path with \o on the curve 
another possible physical
path results. For \o on the path, $x^2$ is always
positive and $S^2$ in Eq~[\ref{trfa}] cannot change sign and all
relatively
time-like pairs of points remain time-like. This operation is like
configuration 4 of the euclidean problem where, according to the
table,  $n$ is inversion invariant.
It might be interesting  to investigate if the inversion
method could be helpful in analyzing certain practical radiation
problems.

\section*{Acknowledgments}
 I am grateful to David Gross for stressing the interest of
studying   the inversion, for
several discussions, and for bringing the
references \cite{gd} and \cite{mf} to my attention.
I would also like to thank E. De Rafael for conversations
concerning
dilogarithms and related problems,  as well as E. Seiler  for
discussions and a  reading of the manuscript.



\begin{references}
\bibitem{line} We assume in our treatment that $C$ becomes
identically the straight line beyond some point. Presumably in a
more sophisticated analysis this could be replaced by a
sufficiently rapid approach to a straight line, to define more
precisely the class of curves included in the argument.
\bibitem{pairs}An
examination of the resulting paths with the associated crossings of
the light
cone might suggest an  interpretation in terms of pairs of
particles.
\end{references}
\end{document}